\documentclass[12pt, preprint]{aastex}

\begin{document}

\title{Measuring Extinction Curves of Lensing Galaxies}

\author{Christina McGough\altaffilmark{1}, Geoffrey
C. Clayton\altaffilmark{1}, Karl D. Gordon\altaffilmark{2}, \& Michael
J. Wolff\altaffilmark{3}}

\altaffiltext{1}{Department of Physics \& Astronomy, Louisiana State
   University, Baton Rouge, LA 70803; Email:cmcgou1@lsu.edu,
   gclayton@fenway.phys.lsu.edu}
\altaffiltext{2}{Steward Observatory, University of Arizona, Tucson,
   AZ 85721; E-mail: kgordon@as.arizona.edu} 
\altaffiltext{3}{Space Science Institute, 3100 Marine Street, Ste
   A353, Boulder, CO 80303-1058; Email: wolff@colorado.edu}

\shorttitle{Extinction Curves of Lensing Galaxies}
\shortauthors{McGough et al.}

\begin{abstract}
We critique the method of constructing extinction curves of lensing
galaxies using multiply imaged QSOs.  If one of the two QSO images is
lightly reddened or if the dust along both sightlines has the same
properties then the method works well and produces an extinction
curve for the lensing galaxy.  These cases are likely rare and
hard to confirm.  However, if the dust along each sightline has
different properties then the resulting curve is no longer a
measurement of extinction.  Instead, it is a measurement of
the difference between two extinction curves.  This ``lens difference
curve'' does contain information about the dust properties, but
extracting a meaningful extinction curve is not possible without
additional, currently unknown information.  As a quantitative
example, we show that the combination of two Cardelli, Clayton, \&
Mathis (CCM) type extinction curves having different values of $R_V$
will produce a CCM extinction curve with a value of $R_V$ which
is dependent on the individual $R_V$ values and the ratio of V band
extinctions.  The resulting lens difference curve is not an
average of the dust along the two sightlines.  We find that lens
difference curves with any value of $R_V$, even negative values, can
be produced by a combination of two reddened sightlines with
different CCM extinction curves with $R_V$ values consistent with
Milky Way dust ($2.1 \leq R_V \leq 5.6$).  This may explain extreme
values of $R_V$ inferred by this method in previous studies.  But
lens difference curves with more normal values of $R_V$ are just as
likely to be composed of two dust extinction curves with $R_V$
values different than that of the lens difference curve.  While
it is not possible to determine the individual extinction curves
making up a lens difference curve, there is information about a
galaxy's dust contained in the lens difference curves.  If the lens
difference curve can be fit with the CCM relationship (regardless of
the fitted $R_V$ value), this implies that the dust along the two
slightlines can be described by CCM.  In addition, the presence of the
2175~\AA\ feature in the lens difference curve means that this feature
is present along at least one of the two lensed sightlines.
\end{abstract}


\section{Introduction}
The well-known Milky Way extinction relation of Cardelli, Clayton, \&
Mathis (1989, CCM) applies to a wide variety of Galactic interstellar
environments.  CCM extinction curves are described by one parameter,
$R_{V}$, the ratio of total-to-selective extinction in the V band.
The $R_V$ value is a rough measure of average dust grain size, and
therefore provides a physical basis for the variations in extinction
curves.  For the Galaxy, the often-quoted average \(R_{V} \simeq 3.1
\), merely reflects the typical value for dust along diffuse
sightlines. But the diverse environments within the Milky Way show a
range of values, \( 2.1 \leq R_{V} \leq 5.6 \) (Valencic, Clayton, \&
Gordon 2004).  A Milky Way UV extinction curve is characterized by a
``bump'' at 2175 \AA\ and a rise in the far-UV.  While CCM applies to
most sightlines in the Milky Way and a few in the Magellanic Clouds,
it does not typically apply in other extragalactic environments
where extinction curves have been measured
(Gordon et al. 2003; Gordon, Calzetti, \& Witt 1997; Pitman, Clayton,
\& Gordon 2000).

Traditionally, the pair method is used to determine extinction by
comparing the spectral energy distributions (SEDs) from reddened and
unreddened stars of the same spectral type (Massa, Savage, \&
Fitzpatrick 1983).  The pair method has only been applied within Local
Group galaxies such as the Milky Way, SMC, LMC, and M31 because
it is necessary 
to use individual stars as point sources.  Determining extinction in
galaxies beyond the Local Group is generally a more complex problem.
Any surface photometry of a galaxy contains light from a mixture of
gas, dust, and stars, making it difficult to analyze the spectral
energy distributions from these complex systems directly.  To derive
the SED of a galaxy requires a radiative transfer model which accounts
for the physical properties of the dust, the stellar population, and
the geometrical distribution of the dust within the galaxy (e.g.,
Gordon et al. 2001; Misselt et al. 2001).  Applied to starburst
galaxies, these models imply that these galaxies have dust with
properties more similar to that in the SMC than to the Galaxy, having
no measurable 2175 \AA\ bump (Calzetti, Kinney, \& Storchi-Bergmann
1994; Gordon et al. 1997).  The existence of the CCM relation in the
Galaxy, valid over a large wavelength interval, suggests that the
environmental processes that modify the grains are efficient and
affect all grains.  In other galaxies, environmental processing by
radiation and shocks near regions of star formation or variations in
galactic metallicity may be responsible for the large observed
variations in extinction curve properties from CCM (e.g., O'Donnell \&
Mathis 1997; Gordon \& Clayton 1998).  Certainly, the application of
the CCM extinction law for Galactic dust is not appropriate for dust
in starburst galaxies and it is unkown if it is appropriate for dust
in more quiescent galaxies.

In order to study extinction in galaxies beyond the Local Group, a new
method has been developed.  It uses multiply imaged QSOs produced by
gravitational lensing to study the extinction properties of the
lensing galaxy (Nadeau et al. 1991).  The multiple images originate
from the same QSO and thus have the same intrinsic SED.  Therefore,
this seems ideal for constructing extinction curves by the pair
method.  The different images are produced by light traveling along
different sightlines through the lensing galaxy, so the amount of
reddening varies from image to image.  The SEDs of the two images are
then compared using the pair method to determine an extinction curve
for the lensing galaxy.  In this paper, we investigate the usefulness
of this method in producing extinction curves of lensing galaxies.

\section{Background}

The literature currently contains $\sim$25 multiply imaged QSOs,
listed in \emph{Table 1}, for which extinction curves have been
calculated, using either spectroscopic or photometric data.
Typically, a CCM relation was fit to these curves and a value of $R_V$
found.  Gravitational lensing is an achromatic process, but often the
various images of a QSO are found to have different colors.  The color
differences are likely due to reddening by dust in the lensing galaxy
(Yee 1988). Each image produced by a gravitational lens is also
magnified by a different amount depending on the geometry of the
source, and the lens, and possibly amplified due to microlensing
effects.  Early work by Nadeau et al. (1991) on the lens system
Q2237+0305 showed that each of four images from a single QSO source
had different colors, and a method for determining the extinction was
suggested. In order to find \(A(\lambda)/A(V)\), the ratio of total
extinction at a given wavelength, $\lambda$, to the total extinction
in the V-band, directly from the ratio of fluxes from two images, it
is necessary to know the amplification ratio.  Since the amplification
ratios are uncertain, they assumed a value of \(A(\lambda)/A(V)\) at
the central wavelength of the K-band.  Because the images only
differed in magnitude by a small amount, Nadeau et al. assumed that
the extinction at any given wavelength differs only slightly from the
average Galactic extinction law.  Nadeau et al. (1991) determined an
extinction law for the lensing galaxy using the assumption that
\(A(K)/A(V)=0.11\), the Galactic value (Clayton \& Mathis 1988).  The
resulting extinction curve is remarkably similar to the $R_V=3.1$
Galactic extinction curve, for $0.5\lesssim \lambda^{-1} \lesssim
2\,\micron^{-1}$.

Falco et al. (1999) built upon the work by Nadeau et al. (1991) and
completed the first survey of extinction properties of 23 lensing
galaxies.  Using photometric data for multiply imaged QSOs, Falco et
al. determined the differential extinction, \(\Delta E(B-V)\), for
each set of images.  The process of determining differential
extinction depends on several assumptions.  First, they assumed that
the source spectrum (from the QSO) is identical for each image.
Second, they assumed that the spectrum does not show significant time
variability over a span equal to the time delay for the images.
Finally, they assumed that the magnification, $\Delta M$, by the lens
does not depend on wavelength or time.  In order to compare the
images, the magnification of one image was set to one, and the
extinction of the bluest image set to zero.  In several cases, Falco
et al. use radio data to determine $\Delta M$.  They use some
spectroscopic redshifts, but also use a technique suggested by Jean \&
Surdej (1998) to determine photometric redshifts for several lenses.
It should be noted that microlensing effects are not considered here,
but Falco et al. suggested that these effects are small compared to
the mean magnification, which is constant and achromatic.  In many
cases for which radio data were unavailable, the differential
magnification, $\Delta M$, was treated as a free parameter when
determining the best fit extinction curve.  Extinction curves have
also been calculated from multiply imaged QSOs by Toft, Hjorth, \&
Burud (2000), Motta et al. (2002), and Munoz et al. (2004).

Motta et al. (2002) use spectroscopic data to determine differential
extinction in the lens galaxy SBS 0909+532.  They suggested that
spectroscopy is more suitable for the study of extinction in lenses
because broadband photometry sometimes makes extinction and
gravitational microlensing indistinguishable. Microlensing causes the
flux for a single image to increase inhomogeneously across the
wavelength range of an individual filter, affecting the color of the
individual image.  Using spectroscopy allowed them to overcome many of
the challenges presented by photometric methods.  Wucknitz et
al. (2003) used spectra of the doubly imaged QSO HE 0512-3329 to
distinguish between the effects of extinction and microlensing.
Because extinction affects the emission lines, but microlensing does
not, the emission lines can be used to separate these effects.  Then,
the continuum, which is affected by both microlensing and extinction,
can be corrected for extinction and the microlensing effects may be
studied directly.  While extinction curves were not the focus of this
paper, they fit the CCM relation to their data.  They treated $R_V$ as
a free parameter, and fit their data using several values of $\Delta
E(B-V)$ and $\Delta A(V)$.  They suggested that the measured
extinction curve could be the result of dust with two different values
of $R_V$ along the two sightlines.

Each of the papers cited above uses roughly the same method for
obtaining an extinction curve by the pair method. When calculating the
extinction curve, they consider $R_V$ to be a free parameter as well
as the magnification, if it is not known.  For several of the lensing
systems, observations in the radio, where extinction is negligible,
allow for a direct measurement of the differential magnification
($\Delta M$).  The magnitudes of two images, labeled A and B, at
wavelength $\lambda$, are compared using the equation
\begin{eqnarray}
 m_{B}(\lambda) - m_{A}(\lambda) = \Delta M +\Delta E(\lambda) R(\frac{\lambda}{1+z_{l}}) 
\end{eqnarray}
where $\Delta E$ is the differential reddening, and $R$ is the ratio
of total-to-selective extinction for the given wavelength and lensing
galaxy redshift (Motta et al. 2002).  This is similar to the
traditional pair method where \(m_{B}(\lambda) - m_{A}(\lambda) =
\Delta E(\lambda) R(\lambda) = A(\lambda)\).  The lensing galaxy
extinction curves are typically presented unnormalized making it
difficult to compare these results to other extinction studies.
Extinction curves should to be normalized to $E(B-V)$ or $A(V)$ so that
only the wavelength dependence, not the amount of extinction, is
compared.  In this paper, all plots are presented in rest wavelength
space and with the extinction normalized to $A(V)$.  It is
straightforward to determine the normalized extinction using the
following equation
\begin{eqnarray}
 \frac{A(\lambda)}{A(V)} = \frac{ m_{B}(\lambda) - m_{A}(\lambda) - \Delta M}{\Delta E(B-V) R_{V}}
\end{eqnarray}

\section{Discussion}

Figure 1 shows a normalized extinction curve for one of the lenses,
B1600+434 (Falco et al. 1999).  Using the photometric data in Falco et
al., as well as their values for $R_V$, $\Delta E(B-V)$, and redshift,
we plot the data and the fit which has an extremely low $R_V = 0.92$.
As pointed out by Falco et al. and shown in the figure, the small
number of photometric points covering a limited wavelength range can
be fit with CCM curves showing a very large range of $R_V$ values.

The results for the extinction curve given in Motta et al. (2002) for
SBS 0909+532 differ greatly from those given by Falco et al. (1999)
for the same QSO.  In fact, the differential extinction is similar
(both reported values around \(\Delta E(B-V) = 0.2\)) but the inferred
values for $R_{V}$ are different.  Falco et al. (1999) used an
estimated redshift of $z = 0.60$ for the lensing system and get $R_V =
0.64 \pm 0.15$.  Their data and the CCM extinction curve with the
reported $R_V$ value are shown in Figure 2.  Motta et al. (2002) used
a lens redshift of $z = 0.83$, found spectroscopically by Oscoz et
al. (1997) using the Mg II absorption doublet and found $R_V = 2.1 \pm
0.9$.  If the extinction curve data given in Falco et al. (1999) are
transformed to the same redshift as Motta et al. (2002) and normalized
with the proper $A(V)$, the data fall right on the curve suggested by
Motta as seen in Figure 2.  Falco et al. (1999) point out that using
the technique of Jean \& Surdej (1998) does not always lead to
accurate redshift and that using the wrong redshift will produce an
incorrect extinction curve.  Therefore, extinction curves produced
from data with uncertain redshifts are not necessarily useful.

Two important assumptions, made by all the studies listed in Table 1,
are that one sightline is less reddened than the other, and that both
sightlines have dust with the same extinction properties, i.e., the
same value of $R_V$.  If both images of the QSO are reddened but with
different values of $R_V$ then combining them to produce an extinction
curve is not so straightforward.  The wavelength dependence of dust
extinction in any galaxy is unknown a priori, but for simplicity let's
assume it follows the CCM relation.

The CCM relationship  at any wavelength, $\lambda$, is:
\begin{equation}
\frac{A(\lambda)}{A(V)} = a(\lambda) + \frac{b(\lambda)}{R_V}.
\end{equation}
Let's assume that each QSO image has a dust column with different
amounts and $R_V$ values.  Then, the magnitude
difference between two lensed images is:
\begin{eqnarray}
A(\lambda) &=& \left[\frac{A_2(\lambda)}{A_2(V)}\right] A_2(V) - 
                 \left[\frac{A_1(\lambda)}{A_1(V)}\right] A_1(V) \\
  &=& \left[ a(\lambda) + \frac{b(\lambda)}{R_{V2}}\right] A_2(V) - 
        \left[a(\lambda) + \frac{b(\lambda)}{R_{V1}}\right] A_1(V) \\
  &=& \left[A_2(V) - A_1(V)\right]a(\lambda) + \left[\frac{A_2(V)}{R_{V2}} - 
        \frac{A_1(V)}{R_{V1}}\right] b(\lambda)
\end{eqnarray}
where $R_{V1}$ and $A_1(V)$ give the dust properties of the least
attenuated, ``unreddened'', image and $R_{V2}$ and $A_2(V)$ of the most
attenuated, ``reddened'', image (Wucknitz et al. 2003).  Normalizing the
extinction at any wavelength by the amount of extinction in the V-band
gives:
\begin{equation}
\frac{A(\lambda)}{A_2(V) - A_1(V)} 
    =  a(\lambda) +
     \left[\frac{A_2(V)}{A_2(V) - A_1(V)}\frac{1}{R_{V2}} - 
            \frac{A_1(V)}{A_2(V) - A_1(V)}\frac{1}{R_{V1}}\right]
        b(\lambda). \\ 
\end{equation}
\emph{Equation 7} gives the extinction curve that is actually measured
for our simple example.  It is the CCM relationship with an $R_V$
value of
\begin{eqnarray}
\frac{1}{R_{V12}} &=& \frac{A_2(V)}{A_2(V) - A_1(V)}\frac{1}{R_{V2}} - 
          \frac{A_1(V)}{A_2(V) - A_1(V)}\frac{1}{R_{V1}} \\
  &=& \left[\frac{1}{1 - \frac{A_1(V)}{A_2(V)}}\right] \frac{1}{R_{V2}} - 
     \left[\frac{1}{\frac{A_2(V)}{A_1(V)} - 1}\right]\frac{1}{R_{V1}}
\end{eqnarray}
Therefore, the combination of two CCM extinction curves is also a CCM
extinction curve but the resulting value of $R_V$ is not an average of
$R_{V1}$ and $R_{V2}$.  Thus, the measurement of the difference
between two lensed sightlines is actually the difference of two
extinction curves, not an extinction curve itself.  In order to avoid
confusion, we propose the term ``lens difference curve'' be used for
these measurements instead of labeling them extinction curves which
can be misleading.

As a quantitative example, in Figure 3 we show the
lens difference curve for B0218+357, described by Munoz et
al. (2004) with 
$R_V = 12$, fit to 6 photometric data points. We have reproduced this
lens difference curve with a combination of different dust types for each
image having $R_V = 5.5$ and $2.7$.  This result is not in any way
unique.  In fact, a very wide range of values of $R_{V12}$, can be
produced by varying the values of $R_{V1}$ and $R_{V2}$.  An example
is shown in Figure 4, with $A_1(V)/A_2(V)=3$.  For this choice of
$A_1(V)/A_2(V)$, we find \(1.5 \lesssim R_{V12} \lesssim 56\). Similar
results are seen for other values of $A_1$(V)/$A_2$(V).  It is even
possible to produce negative values of $R_V$ for such lens difference
curves (Wucknitz et al. 2003).

In general, for any two QSO images, the value of $A_1$(V)/$A_2$(V) is
unknown. This highlights the main difference between the pair method
as applied to lensing galaxies and the pair method as applied to stars
in the Galaxy and the Magellanic Clouds (Valencic et al. 2004; Gordon
et al. 2003). In the latter case, the reddening, E(B-V), and the ratio
of total-to-selective extinction, $R_V$, can be independently measured
along the sightline to each star so that $A_1$(V)/$A_2$(V) is a known
quantity. Then, a unique extinction curve, within the measurement
uncertainties, can be derived. Only in two limiting cases will the
pair method 
produce an actual extinction curve for QSO images. First, if
the wavelength dependence of the extinction is the same for dust along
both sightlines then the resulting lens difference curve will be
an extinction curve. In our simple CCM example, if the value of
$R_V$ is the same 
for the dust along both sightlines, then equations 7 and 9 produce a
correct 
extinction curve and $R_V$ value. Second, if $A_2$(V) $>>$
$A_1$(V) then it doesn't matter if the extinction parameters are
different for the dust along the two sightlines. In this case,
equations 7 and 9 again produce the correct values when $A_1$(V) is
set to zero.

So the pair method for lensing galaxies can work but generally it is
unknown which individual pairs of QSO images either have the same
$R_V$, or have one image that is much more reddened than the other.
Several hundred sightlines within the Milky Way have measured
extinction curves (Valencic et al. 2004).  The so-called average Milky
Way extinction curve with $R_V$=3.1 is, in fact, the typical
extinction for diffuse sightlines in the Galaxy. It is not in any way
an average curve for the Galaxy. Any two sightlines widely spaced in
the Galaxy are likely to have very different values of $R_V$.  The
lensed QSO, B0218+357, where one sightline passes through a molecular
cloud, is a case in point (Munoz et al. 2004).  If the second
sightline is lightly reddened then an accurate extinction curve for
the dust in this molecular cloud is being produced. But if the second
sightline is also significantly reddened but with dust from the more
diffuse interstellar medium in the intervening galaxy, then the
resulting measurement is a lens difference curve not an
extinction curve.  This lens difference curve reflects the difference
in the two sightlines' extinction curves and it is nontrivial to
extract the original extinction curves from the lens difference
curve.  A very wide range of $R_V$ values is listed for
the lensing galaxies in Table 1. Some fall in the normal range seen
for Galactic sightlines, between $2.1$ and $5.6$, while other values
are quite extreme. However, as can be seen from Figure 4, the normal
Galactic values in Table 1 are no more likely to be correct than the
extreme values.  

While it is not possible to extract the extinction curves from a
lens difference curve, there is information about the dust in the
lensing galaxy contained in the lens difference curve.  While it
may be possible to create a lens difference curve which follows the
CCM relationship without the two individual extinction curves also
following the CCM relationship, this seems unlikely.  Therefore, if
the lens difference curve can be fit with the CCM relationship
regardless of the $R_V$ value, this is evidence that the dust in the
lensing galaxy follows the CCM relationship.  In addition, if the
2175~\AA\ extinction bump is present in the lens difference
curve, then there must be bump dust in the more reddened sightline
at least.  Munoz et al. (2004) focus on the extinction bump at
2175~$\AA$.  Lens difference curves are shown for LBQS 1009-0252
and B0218+357, and the latter shows a bump at 2175~\AA\ which is
reproduced here in Figure 3.  Motta et al. (2002) also found the bump
in their lens difference curve for SBS 0909+532.  The presence
of the bump in these lens difference curves implies that there
is Milky Way-type bump dust, not just SMC-type dust along at least one
of the component sightlines.

Just because we do not see very large or very small values of $R_V$ in
the Local Group galaxies doesn't mean that exotic dust with very
different properties can't exist in distant galaxies. Very
simplistically, a small $R_V$ implies that a greater mass fraction is
contained in small dust grains and vice versa.  However, extinction by
large grains is less efficient than extinction by small grains because
mass scales as $a^3$ while surface area scales as $a^2$.  Therefore,
for any given gas abundance of grain materials and gas-to-dust-ratio,
large grains will use a much higher fraction of the available material
in order to produce the same amount of extinction.  For instance
models for dust along Galactic sightlines having high $R_V$, such as
the Orion sightline HD 37022 ($R_V=5.5$), tend to run out of material
to build large grains (Clayton et al. 2003).  Therefore in general,
sightlines with very small values of $R_V$ are more likely to be
physically possible than very large $R_V$ values.

\section{Conclusions}

\begin{itemize}

\item In general, the difference between two lensed sightlines in
a galaxy does not produce an extinction curve, but a lens difference
curve.  It is not easy to decompose this lens difference curve into
the two sightline extinction curves.

\item We show that the combination of two CCM type extinction
curves having different values of $R_V$ will produce a lens difference
curve which follows the CCM relationship.  The $R_V$ value of the lens
difference curve is a function of the $R_V$ values of the individual
lensed sightlines and the ratio of their V band attenuations.

\item If one of the two QSO images is lightly reddened or if the dust
along both sightlines has the same extinction wavelength dependence
then the lens difference curve is an accurate extinction curve
for the lensing galaxy.

\item Lens difference curves can only be accurately constructed
for lenses with known redshift and magnification ratio, $\Delta M$.  If the
redshift is wrong, the lens difference curve is wrong.

\item A small wavelength coverage results in lens difference
curve CCM fits that are not well constrained.

\item The presence of a ``bump'' at 2175~\AA\ in a lens
difference curve implies that there is Milky Way-like dust along at 
least one of the sightlines.

\item An accurate CCM fit to a lens difference curve, regardless
of the fitted $R_V$ values, implies that the two lensed sightlines
have CCM-like dust extinction curves.

\end{itemize}

\begin{deluxetable}{lccccc}
\tabletypesize{\small}
\tablewidth{0pt}
\tablecaption{Summary of Recent Work}
\tablenum{1}
\tablecolumns{6}
\tablehead{ \colhead{object} & \colhead{$R_V$} & \colhead{$\Delta E(B-V)$} & \colhead{z\tablenotemark{a}} & \colhead{$\Delta M$\tablenotemark{b}} & \colhead{Reference\tablenotemark{c}} \\ 
 \colhead{} & \colhead{} & \colhead{(mag)} & \colhead{} & \colhead{} & \colhead{} } 
\startdata
    Q0142-100 & $3.11 \pm 1$ & -0.06 & 0.49 & \nodata  &   1 \\ \hline
    B0218+357 & $7.2 \pm 0.08$ & 0.62 & 0.68 & $1.4 \pm 0.03$ &   1  \\ 
         & $12 \pm 2$ & 0.30 &   0.6847 & $1.4 \pm 0.3$ &  2 \\ \hline 
    MG 0414+0534 & $1.47 \pm 0.15$ & 1.41 & 0.96 &   &   1 \\ 
         & 1.8 & \nodata  & (1.15) & \nodata &   3 \\ \hline 
    SBS 0909+532 & $0.64 \pm 0.15$ & 0.19 & (0.60) & \nodata &   1 \\ 
         & $2.1 \pm 0.9$ \tablenotemark{d} & 0.20 &   0.83 & $-0.2 \pm 0.2$ \tablenotemark{e} & 4 \\ \hline 
    FBQ 0951+2635 & $4.86 \pm 0.85$ & -0.03 & (0.30) & $1.67 \pm 0.22$ &   1 \\ \hline
    BRI 0952-0115 & $3.1 \pm 1$ & 0.12 & (0.55) & \nodata &  1  \\ \hline
    Q0957+561 & $6.63 \pm 0.87$ & -0.05 & 0.36 & \nodata &  1 \\ \hline
    LBQS 1009-0252 & $2.72 \pm 0.84$ & -0.14 & (0.60) & \nodata &   1 \\ 
          & $2.5 \pm 0.3$ & 0.41 &   (0.88) & $1.3 \pm 0.1$ &   2 \\ \hline 
    Q1017-207=J03 & $3.1 \pm 1$ & -0.03 & (0.60) & \nodata &   1 \\ \hline
    B1030+074 & $3.1 \pm 1$ & 0.38 & 0.6 & $0.32 \pm 0.04$ &  1 \\ \hline
    HE 1104-1805 & $2.87 \pm 0.96$ & -0.14 & (0.80) & \nodata &   1 \\ \hline
    PG 1115+080 & $2.89 \pm 0.99$ & -0.02 & 0.31 & \nodata &   1 \\ \hline
    Q1208+1011 & $3.47 \pm 0.97$ & 0.07 & (0.60) & \nodata &   1 \\ \hline
    HST 12531-2914 & $3.1 \pm 1$ & \nodata  & (0.81) & \nodata &   1 \\ \hline
    H1413+117 & $2.94 \pm 0.66$ & 0.22 & (0.70) & \nodata   &   1 \\
             & 9   & \nodata  & (1.15) & \nodata &   3 \\ \hline
    HST14176+5226 & $3.1 \pm 1$ & \nodata  & 0.81 & \nodata &   1 \\ \hline
    B1422+231 & $2.91 \pm 0.81$ & 0.35 & 0.34 & \nodata    &   1 \\ \hline
    SBS 1520+530 & $2.83 \pm 0.96$ & -0.19 & (0.49) & \nodata &  1 \\ \hline
    B1600+434 & $0.92 \pm 0.58$ & 0.22 & 0.42 & $0.29 \pm 0.1$ &  1 \\ \hline
    PKS 1830-211 & $6.34 \pm 0.16$ & 0.57 & 0.89 & $0.46 \pm 0.06$ & 1 \\ \hline
    MG 2016+112 & $2.8 \pm 0.78$ & -0.01 & 1.01 & $-0.06 \pm 0.06$ &  1 \\ \hline
    HE 2149-2745 & $3.08 \pm 1$ & -0.07 & (0.49) & \nodata &  1 \\ \hline
    Q2237+0305 & $5.29 \pm 0.82$ & 0.13 & 0.04 &  \nodata   &  1 \\ 
           & 2.8   & \nodata  & (0.13) & \nodata &   3 \\ \hline
    B1152+199 & $1.8 \pm 0.4$ & 1  & 0.44 & 1.21 &   5 \\  \hline
\enddata

\tablenotetext{a}{parenthesis indicate estimated redshifts with no spectroscopic confirmation}
\tablenotetext{b}{only available for objects with radio data}
\tablenotetext{c}{References.- (1) Falco et al. (1999); (2) Munoz et al. (2004); (3) Jean \& Surdej (1998); (4) Motta et al. (2002); (5) Toft et al. (2000)}
\tablenotetext{d}{Motta et al. use spectroscopic data}
\tablenotetext{e}{$\Delta M$ determined from the best-fit extinction curve}

\end{deluxetable}

\begin{figure}[p]
\centering
\plotone{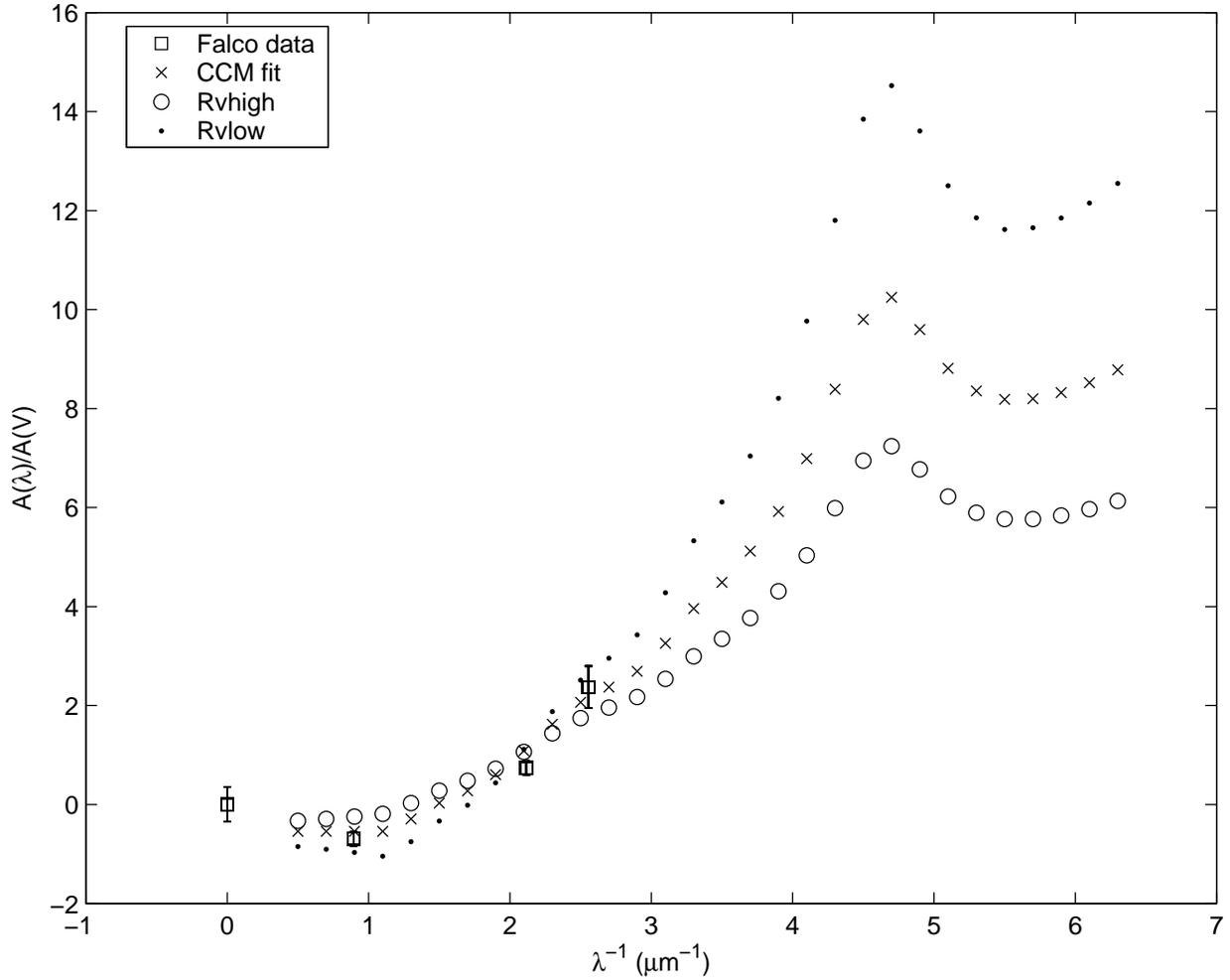}
\caption{Extinction data for B1600+434 (squares) fitted with a CCM
curve with $R_V = 0.92$ (crosses) from Falco et al. (1999).
Additional CCM extinction curves are plotted with Rvlow=0.65 and
Rvhigh=1.3.  This plot shows how weakly constrained the ``fitted'' CCM
extinction curve is.  The data are presented in rest wavelength
space.}
\end{figure}

\begin{figure}[p]
\centering
\plotone{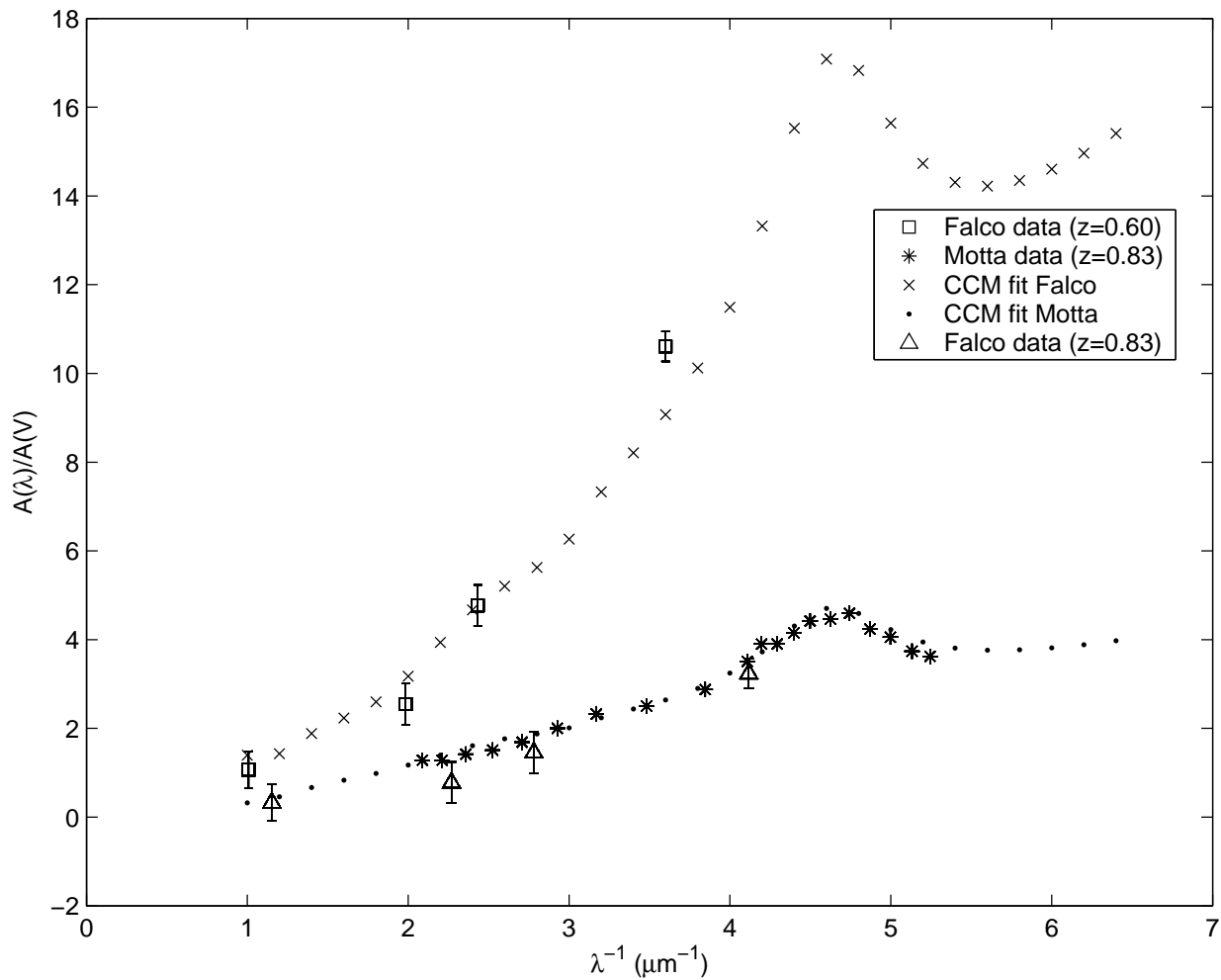}
\caption{Motta et al.(2002) and Falco et al.(1999) data for SBS
0909+532 shown with CCM fits for $R_v = 2.1$, the value obtained in
the fit by Motta et al., and $R_V = 0.64$, the value reported by Falco
et al. for their best-fit extinction curve. Also plotted are the Falco
data using the Motta redshift and normalization, $A(V)$.  All data
have been transformed into rest wavelength space using the reported
redshifts.}
\end{figure}

\begin{figure}[p]
\centering
\plotone{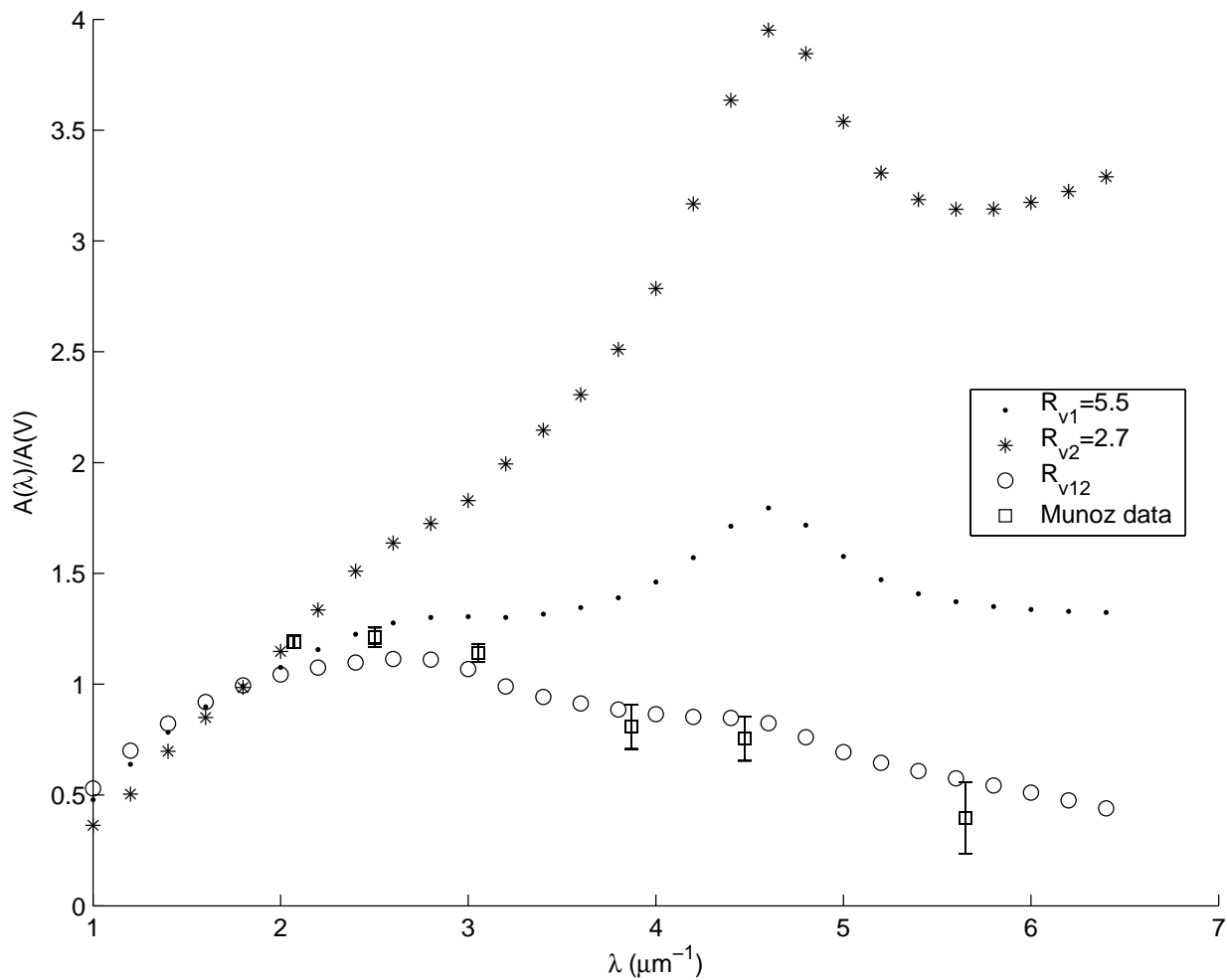}
\caption{Plot includes data from Munoz et al.(2004) for B0218+357,
which fits a CCM ($R_V= 12$) extinction curve to photometric data in
rest wavelength space.  This plot shows two extinction curves with the
following parameters: $R_{V1} = 5.5$, $R_{V2} = 2.7$, $E(B-V)_1= 0.7$,
$E(B-V)_2 = 0.4$ as well as the combined curve that would result which
has $R_{V12} \approx 12$. Note that the choice of $R_{V1}$ and
$R_{V2}$ are not unique.  See text.}
\end{figure}

\begin{figure}[p]
\centering
\plotone{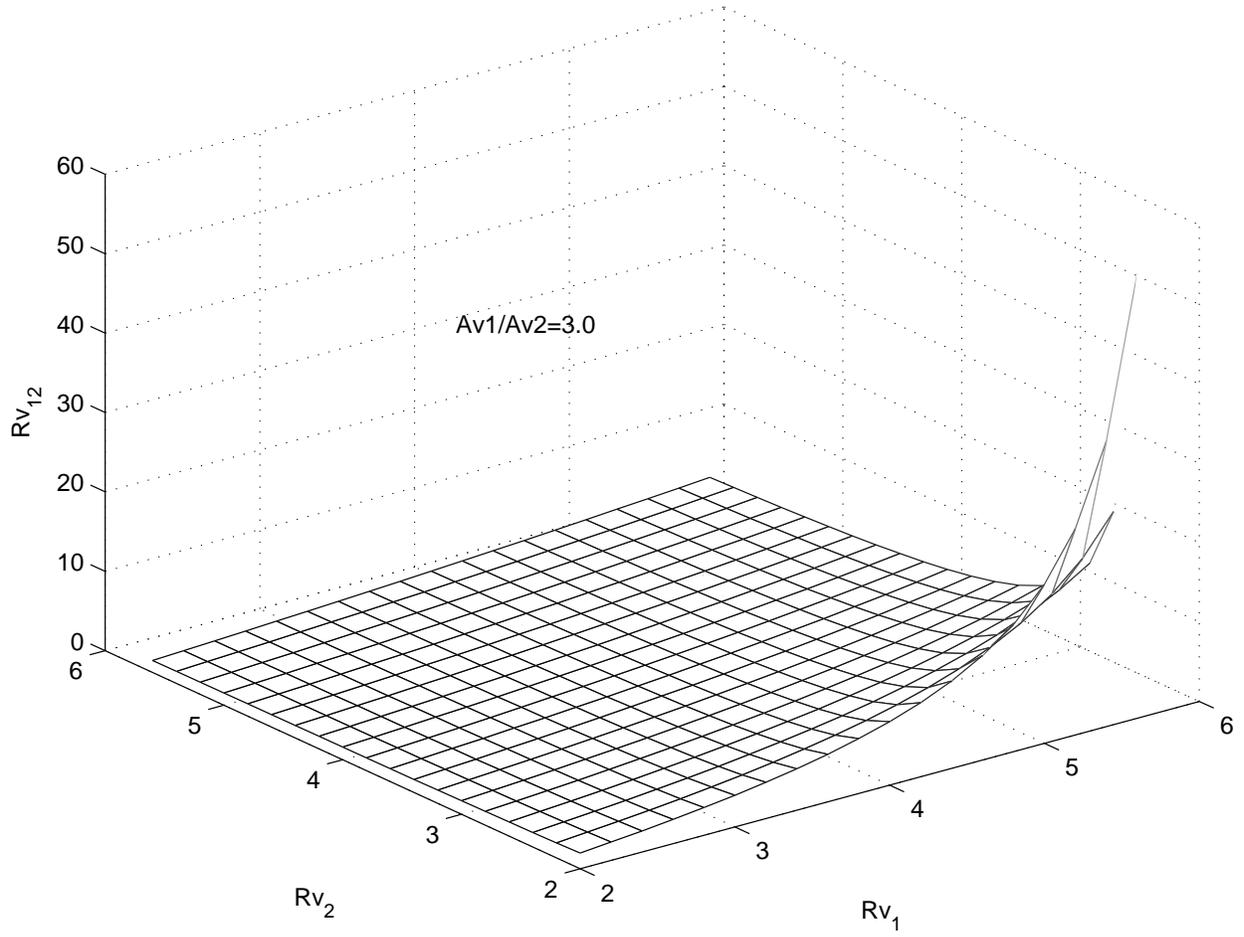}
\caption{For any fixed ratio, $A_1(V)/A_2(V)$, it is possible to
obtain a wide range of combined $R_V$ values, $R_{V12}$.  For this
fixed value of $A_1(V)/A_2(V)=3$, we find \(1.5 \lesssim R_{V12}
\lesssim 56\).}
\end{figure}

\end{document}